\documentclass[journal]{IEEEtran}
\usepackage{cite}
\usepackage{amsmath,amssymb,amsfonts}
\usepackage{algorithmic}
\usepackage{graphicx}
\usepackage{textcomp}
\def\BibTeX{{\rm B\kern-.05em{\sc i\kern-.025em b}\kern-.08em
    T\kern-.1667em\lower.7ex\hbox{E}\kern-.125emX}}
\begin{document}

\title{Proposal-based Few-shot Sound Event Detection for Speech and Environmental Sounds with Perceivers}

\author{\IEEEauthorblockN{Piper Wolters\IEEEauthorrefmark{1}, Logan Sizemore\IEEEauthorrefmark{2}, Chris Daw\IEEEauthorrefmark{3}, Brian Hutchinson\IEEEauthorrefmark{2}\IEEEauthorrefmark{4}, and Lauren Phillips}\\
\IEEEauthorblockA{\IEEEauthorrefmark{1}Allen Institute for Artificial Intelligence\\
\IEEEauthorrefmark{2}Western Washington University\\
\IEEEauthorrefmark{3}DocuSign, Seattle, WA 98104\\
\IEEEauthorrefmark{4}Pacific Northwest National Laboratory, Richland, WA 99354\\
}
\thanks{This work was funded by the U.S. Government.}}

\maketitle

\begin{abstract}
Many applications involve detecting and localizing specific sound events within long, untrimmed documents, including keyword spotting, medical observation, and bioacoustic monitoring for conservation. Deep learning techniques often set the state-of-the-art for these tasks. However, for some types of events, there is insufficient labeled data to train such models. In this paper, we propose a region proposal-based approach to few-shot sound event detection utilizing the Perceiver architecture. 
Motivated by a lack of suitable benchmark datasets, we generate two new few-shot sound event localization datasets: ``Vox-CASE,'' using clips of celebrity speech as the sound event, and ``ESC-CASE,'' using environmental sound events.
Our highest performing proposed few-shot approaches achieve 0.483 and 0.418 F1-score, respectively, with 5-shot 5-way tasks on these two datasets. These represent relative improvements of 72.5\% and 11.2\% over strong proposal-free few-shot sound event detection baselines.
\end{abstract}

\noindent \begin{small} {\it Keywords: } few-shot, localization, perceiver, sound event detection \end{small}

\section{Introduction} \label{sec:introduction}

\IEEEPARstart{S}{ound} event detection is the task of detecting the start and end of specific acoustic events within a longer audio recording. The events of interest depend on the use case; for example, one might wish to detect coughs for a medical observation \cite{liu14coughs}, gunshots for public safety \cite{morehead19guns}, or animal vocalizations for biomonitoring \cite{zeppelzauer2015towards}.
For most real applications, manual sound event detection simply does not scale, necessitating automated systems. A significant body of work exists on sound event detection and adjacent tasks (e.g., keyword spotting). 

One can organize existing sound event detection methods into two major categories: (1) frame- or window-level models that make predictions over frames or fixed-length windows of frames of the signal, denoted here as {\bf window-level}, and (2) \textbf{proposal-based} models that propose and classify variable length temporal regions of interest.
Both strategies have demonstrated promising results for event detection.

Cakir et al. compare purely-convolutional, purely-recurrent and convolutional recurrent neural network (CRNN) window-level approaches to polyphonic sound event detection~\cite{cakir2017convolutional}. Their work notes a lack of precisely-labeled sound event detection datasets, so they include results on their own synthetic dataset. Lim et al. propose a window-level approach using a CRNN to tackle the DCASE 2017~\cite{mesaros17dcase} rare sound event detection task~\cite{lim17rare}.  On a different DCASE 2017 detection task, 
Lu et al. find that bidirectional gated recurrent units (GRUs) do not outperform CRNNs for polyphonic sound event detection~\cite{lu2017bidirectional}.

In contrast to frame- or window-level detection techniques, proposal-based methods break the task into two or more stages: an initial stage that proposes potential event regions and one or more subsequent stages to refine and classify the proposals.
Pham et al. introduce the concept of \textit{eventness} for region-based sound event detection, similar to \textit{objectness}~\cite{pham2018eventness}, used in object detection. Wang et al. propose utilizing the R-FCN architecture, improving the region proposal network (RPN) portion to specifically propose regions which contain audio events~\cite{wang17rfcn}. Inspired by the proposal-based Faster-RCNN~\cite{ren15faster}, Kao et al. propose a region-based CRNN for the sound event detection~\cite{kao18rcrnn} problem. For a weakly supervised sound event detection task, Kiyokawa et al. use a ResNet model with a self-mask module to improve detection accuracy of their RPN. Hou et al. \cite{hou19foot} adopt an RPN to a keyword spotting task, specifically extending their model to be able to detect multiple keywords.

Unfortunately, in many cases, the machine learning methods these systems rely upon require significant amounts of labeled training data for each event of interest.
In contrast, humans can learn a new task after only a few examples. 
 {\it Few-shot learning} is a subfield of machine learning that aims to develop machine learning models able to accurately model previously unseen classes given only a limited number of examples. Many of the early few-shot learning papers focused specifically on few-shot classification of images \cite{vinyals16match,snell17proto,koch2015siamese,sung18relnet}.
 
Few-shot audio {\it classification}, which does not involve temporally localizing events, has been explored in the literature, with strong results reported for sound classification and speaker identification~\cite{wolters20fs,chou19transient,zhang19graph, shi20meta}. Wang et al.~\cite{wang19centroid} utilize prototypical networks~\cite{snell17proto} to improve speaker identification and verification tasks, for both seen and unseen speakers. Chen et al.~\cite{chen20spoken} investigate utilizing few-shot learning  to the task of spoken word classification, posing it as an $N+M$-way problem, where $N$ and $M$ are the number of new classes and base classes, respectively, by modifying the model-agnostic meta-learning algorithm~\cite{finn2017modelagnostic}.

Some window-based few-shot sound event detection methods and studies exist in the literature, mainly in the related area of keyword spotting.
Wang et al. compares four few-shot, metric-based learning strategies using a convolutional neural network (CNN) embedding module, and validates their approach on a keyword spotting task using Spoken Wikipedia Corpora~\cite{wang20fewsed}. Their work concludes that the prototypical network performs best in this domain. Parnami and Lee \cite{parnami20protowords} use the prototypical network with embeddings from their proposed ResNet-based architecture for the few-shot keyword spotting task, and demonstrate the ability to spot new keywords with few examples.
Outside of few-shot keyword spotting, Shimada et al.~\cite{kaz20background} propose a few-shot sound event detection framework that includes an explicit class for background noise, and they evaluate their approach on the well-annotated DCASE2017 Task 2. Whereas this existing few-shot sound event detection literature uses window-level approaches, 
to the best of our knowledge, we are the first to use a proposal-based model for few-shot audio localization, whose performance we further extend by incorporating the Perceiver architecture.

While most of the models mentioned above utilize CNNs and recurrent neural networks (RNNs), Transformer-based models\cite{vaswani17transformer} are also quite popular for audio tasks \cite{gulati20conformer, jaegle21perceiver}. Self-attention architectures have shown success, not only with text, but with vision, audio, and video \cite{vaswani2017attention, dosovitskiy2020image, baevski2019vq, girdhar2019video}. In the few-shot setting, Ye et al. found success by simply using a self-attention submodule before a few-shot classifier \cite{ye2020few}. Chou et al. \cite{chou19transient} and Zhang et al. \cite{zhang19graph} utilize attention in few-shot audio classification (not localization) tasks. In the sound event detection realm, most methods that incorporate attention are focused on a semi-supervised task, e.g., \cite{koichi20weakly,koichi20conformer}.
Yan et al. \cite{yan19region} propose a region-based attention method to further boost the success of CRNNs on weakly supervised sound event detection. Jaegle et al. \cite{jaegle21perceiver} introduce the modality-agnostic {\it Perceiver} architecture, which scales to very large inputs and achieves impressive results in a variety of modality specific tasks, including audio tasks. 
We use the Perceiver as a submodule in some of our models. 

Lastly, there is a recent body of work developing few-shot event detection techniques for video data, including both (video) frame-level \cite{yang_one_2018,zhang2020metal} and proposal-based \cite{xu_similarity_2018,chao2018rethinking,zhang2020zstad} approaches. We draw particular inspiration for our proposal-based few-shot methods from \cite{xu20video}, which presents a few-shot temporal activity detection framework, based on proposal regression.

The specific contributions of this paper are:
\begin{itemize}
    \item We introduce a proposal-based few-shot audio event detection model following the video event detection model of~\cite{xu20video}. To the best of our knowledge, this is the first proposal-based model reported in the literature for few-shot sound event detection. 
    \item We study the benefits of incorporating attention into our proposal-based localization methods, utilizing the scalable and high-performing Perceiver \cite{jaegle21perceiver} architecture. 
    \item We provide a thorough evaluation of a window-level baseline model and proposal-based models with and without the Perceiver component on the two novel benchmarks suitable for few-shot sound event detection.
\end{itemize}

\section{Few-shot Learning}

Few-shot learning refers to the scenario in which one must make predictions with very limited training data. Whereas traditional machine learning models may require hundreds or thousands of examples of each class, few-shot learning models are given only a few (e.g., 1-10) labeled training examples per class.

\subsection{Episodic Training} \label{ssec:episodic}
Few-shot methods commonly employ {\it episodic training} \cite{vinyals16match}, in which the few-shot learner is trained by solving a series of episodes (tasks). An episode $E = (S, Q)$ consists of a {\it support set} $S$ and a {\it query set} $Q$. The support set contains all of the training data for the task. For example, for a $k$-shot $n$-way task, the support set consists of $n$ classes, each with $k$ examples. After ``training'' on the support set, the few-shot method must make predictions on the query set samples. The overall ``meta-training'' process consists of repeatedly sampling tasks from a set of training classes, and minimizing a suitable loss on each query set.
Note that episodes in few-shot-learning play the role of minibatches often found in non-few-shot deep learning training.
For our sound event detection task, the query set, $Q$, contains untrimmed audio files containing one or more of the events from the episode's $n$ classes, and our support set contains $k$ (disjoint) trimmed examples for each of the $n$ classes. Regions of the query files where no event is taking place are labeled as ``background,'' also referred to as ``no event." 

\subsection{Prototypical Networks}
Prior work has found prototypical networks work well for keyword spotting~\cite{wang20fewsed}, so we adopt them here for few-shot sound event detection. Prototypical networks are a metric-based approach that involves training a neural network $f_\theta$ that embeds datapoints into an embedding space. The model produced by the few-shot learner is itself non-parametric: the $k$ support set embeddings per class are averaged to produce $n$ class prototypes, $c_1,c_2,\dots,c_n$: 
\begin{equation}
    c_m = \frac{1}{|S_m|} \sum_{s_i\in S_m} f_\theta (s_i).
\end{equation}

Here $S_m$ denotes the subset of the support set belonging to class $m$.
For each data point $q$ in $Q$ (e.g., each proposal), the probability distribution $\hat{y}$ over class labels is given by 

\begin{eqnarray}
    \hat{y} & = & \mbox{softmax}(z)\\
    z_n & = & -d(f_\theta (q), c_n),
\end{eqnarray}

where $d$ is a distance function (Euclidean distance in our experiments).

\section{Dataset Generation} \label{sec:datagen}

Based upon code for DCASE 2017 Task 2~\cite{mesaros17dcase}, we generate two new few-shot sound event detection datasets on which we evaluate our models:

\subsubsection{ESC-CASE} \label{sec:ESC-CASE}
ESC-CASE events come from the ESC-50 dataset~\cite{karol15esc}. 
The 50 environmental sound classes are randomly split 30/10/10 for train/validation/test. 
We generate 25000 training episodes, 5000 validation episodes, and 5000 test episodes.

\subsubsection{Vox-CASE}
For Vox-CASE our events are speaker clips from VoxCeleb2~\cite{chung18vox}. Our classes are individual speakers, and we split them into 4417, 1473, and 1473 training, validation, and test classes, respectively. The speaker clips have a wider range of duration than ESC-50, so in order to fit 1-3 speakers in a 30 second clip, we (uniformly) randomly center crop each speaker clip over five seconds long into the range $[2,5]$ seconds.
Like ESC-CASE, we generate 25000 training episodes, 5000 validation episodes and 5000 test episodes.\\

Each training episode contains events from only $n$ classes randomly sampled from the set of training classes; validation and test episodes are created analogously. Each episode's support set contains $k$ trimmed examples of each of $n$ classes. The query set contains eight longer (untrimmed) query clips, each with random background noise overlaid with 1-3 sound events. The background noises are from TUT Acoustic Scenes \cite{mesaros2016tut}. Disjoint sets of background audio are used for training (822 background clips), validation (150 clips), and test (150 clips).
The event-to-background ratio (EBR) \cite{mesaros17dcase} of each overlaid event is uniformly sampled among \{-12, -6, 0, 6, 12\}dB. Varying EBR adds a level of difficulty to the task; in particular, events with low EBR are quieter and much harder to distinguish from the background noise.

\section{Methods}

\subsection{Feature Extraction}

All raw audio is resampled to 16kHz and then converted to 64-bin log-mel spectrograms using a frame length of 25ms and an offset of 10ms. All of the (untrimmed) query set clips are 2998 frames (30 seconds). 
Support set clips are uniformly randomly cropped to be between 1 and 5 seconds for ESC-CASE and 2 and 5 seconds for Vox-CASE.

\begin{figure*}
  \includegraphics[width=1\linewidth]{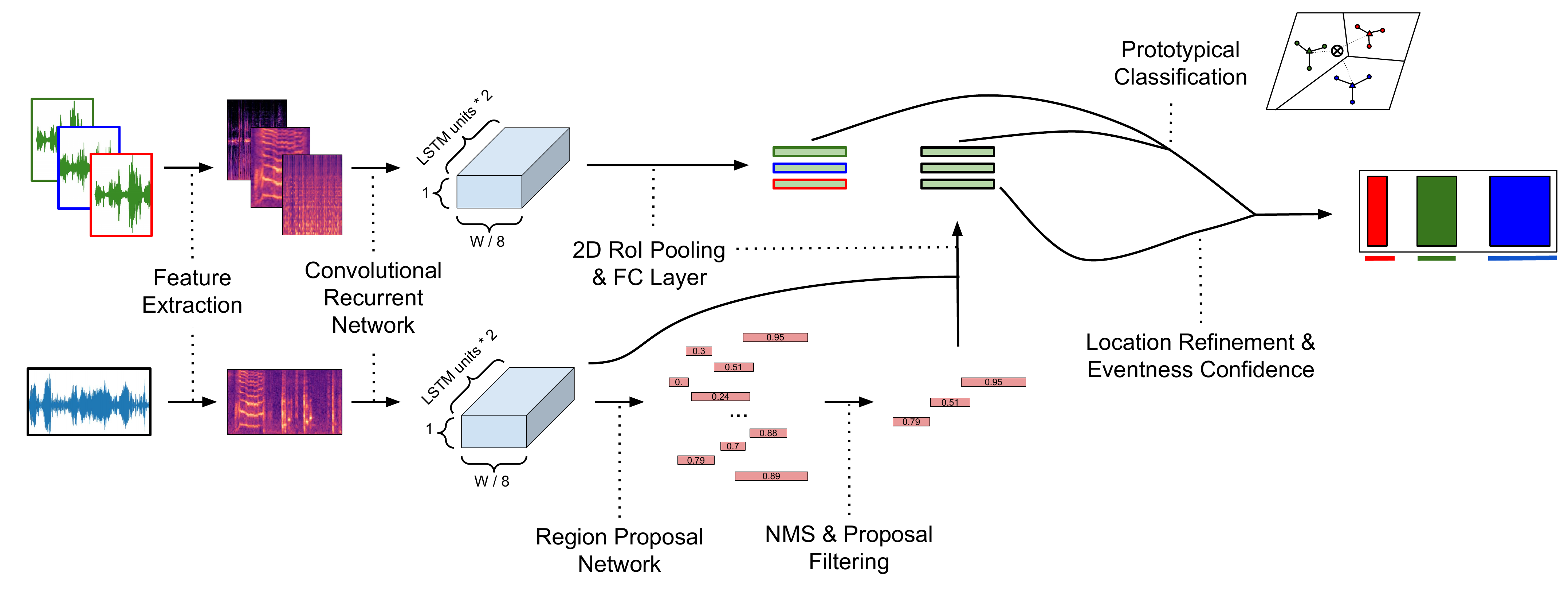}
  \caption{An overview of our proposed Proto-RCRNN architecture. The model has four components: (1) a convolutional recurrent neural network (CRNN) that generates embeddings for both supports and queries, (2) a region proposal network (RPN) that proposes regions of interest, (3) filtering of highly overlapping proposals and region refinements, (4) a prototypical network that classifies each proposal for a given query by computing the distance from prototypes computed using the support embeddings.}
  \label{fig:protoarch}
\end{figure*}

\subsection{Proposal-Based CRNN Model}

Our proposal-based model, denoted the Proto-RCRNN, is illustrated in Fig. \ref{fig:protoarch}. It contains four major subcomponents.

\subsubsection{Convolutional Recurrent Neural Network}
The backbone is a CRNN, based upon \cite{cakir2017convolutional}.
It takes as input the log-mel spectrograms,
for both the support and query sets. The model consists of seven convolutional layers, each followed by Leaky Rectified Linear Units and max pooling layers, followed by a bidirectional Long Short-Term Memory network (LSTM) \cite{lstm}. The result is a sequence of per-timestep feature vectors, with time resolution downsampled by a factor of 8.

\subsubsection{Region Proposal Network}\label{sec:rpn}
Our RPN is based on R-CRNN, another region-based convolutional recurrent neural network by Kao et al.~\cite{kao18rcrnn}, which itself is an adaptation adding  a recurrent network to Faster RCNN~\cite{ren15faster}.
At each step of the sliding window over the feature map, the current window is mapped down to a 512-dimensional feature vector. 
Multiple ``anchor" regions of varying sizes are proposed at the center each of these windows. The base size of the anchors is 4, with anchor scales in $\{1, 2, 4, 8, 16\}$ meaning the actual size of anchors ranges from $4=4\cdot 1$ frames to $64=4\cdot 16$ frames. These anchors are repeated at a stride of $2$ frames. 

The RPN returns $p$ proposals. 
Specifically, the 512-dimensional embedding vector from each location of the feature map is fed to two layers: (1) one dense (binary classification) layer to predict the probability of having an event; this outputs $p$ scores, (2) another dense (regression) layer to generate anchor location refinements. 
For each anchor, the regression layer outputs a length offset, $lo$, and a center offset, $co$.
Given that an unrefined RoI has a length $L$ and a center $C$, the length of the refined RoI will be $L' = L + lo$, and the center of the refined RoI will be $C' = C + co$, thus the [start, end] coordinates will be $[C' - (L' / 2), C' + (L' / 2)]$.

The targets for the classification and the regression layers are decided by the overlap between each unrefined anchor and the ground truth events. During training, we require an intersection-over-union (IoU) threshold of $\geq.7$ in order for an anchor to be given a positive (event) class label; proposals with an IoU between $.3$ and $.7$ are given no label; finally, those with an IoU $\leq.3$ are considered non-events (i.e., background).

For both training and test time, any region proposals that are less than one frame long are removed. If the boundaries of the proposal are outside of the audio's start and stop time, the proposal is clamped to the boundaries. The proposals are then filtered using non-maximum suppression, with a threshold of .1, and the top $s$ scored proposals are passed onto the next layer, encouraging the model to be most confident about regions containing events. Hyper-parameter $s$ is chosen via tuning on the validation set.

\subsubsection{RoI Pooling, Proposal Filtering and Region Refinements}
\label{section:roipool}

Each event proposal from the RPN is fed into the RoI pooling layer, which maps a variable length segment to a fixed-length feature vector (twice the number of hidden units). The output of this pooling layer is fed through a series of dense layers.  The same proposal refinement used after the RPN is also applied after the dense layers.

\subsubsection{Prototypical Classification}
The top $s$ scored, refined proposals from the proposal refinement and filtering layer are passed in to this classification layer. The output of the prototypical network for each proposal represents the confidence scores for each class. Prior to applying the output softmax, we append a learned scalar logit that corresponds to ``no event," allowing us to predict nothing when no class is sufficiently confident.

\subsection{Proto-RCRNN + Perceiver}
\label{section:proto-perceiver}
For our Proto-RCRNN + Perceiver model, we replace the RoI pooling layer, as described in Section \ref{section:roipool}, with a Perceiver \cite{jaegle21perceiver}.  For each of the top $s$ scored proposals, we extract the respective regions out of the feature map, output by the backbone. Each of these variable length regions is fed through the Perceiver to obtain a fixed-length vector, used analogously to how the RoI pooling layer output is used in the previous subsection.

\subsection{Window-level CRNN Baseline}
\label{section:window}
As a window-level baseline, we implement and evaluate a 1d CRNN inspired by~\cite{lim17rare}.
The input to this model is consecutive overlapping windows of spectrogram frames. The windows are 128 frames long overlapped using a 64-frame offset between windows. Every window of the spectrogram is encoded with an embedding network consisting of a CRNN; namely, a 2d 5-layer CNN followed by a 2-layer bidirectional LSTM. Each resulting window-level vector is classified by a prototypical network given the prototypes generated by the support set $S$. We use an extra learned logit in the prototypical network to act as a ``distractor" to classify ``no event" frames if no other logit is sufficiently confident. 

\subsection{Training}\label{sec:training}

We train our proposal-based models to optimize the weighted combination of three loss terms: 
\begin{enumerate}
    \item The RPN loss, which is a weighted combination of binary focal loss \cite{yi17focal} on the classification layer output and smooth L1 loss on the regression layer output. 
    \item The proposal refinement loss, which contains the same two loss terms as the RPN loss, but applied during the proposal refinement stage.
    \item Focal loss for the per-proposal prototypical network classification output.
\end{enumerate}
The classification and regression targets for the first term are determined by overlap (specifically, IoU) between the anchors and ground truth events. The targets for the second loss term is determined analogously, except using (unrefined) RPN proposals instead of anchors.

The use of focal loss is motivated by the large class imbalance between sound event and background: it further down-weights the loss contributions for datapoints on which the model is confident, encouraging the model to focus on challenging datapoints instead of further driving down loss for obvious ``no event'' proposals.  We refer readers to Lin et al. \cite{yi17focal} for additional details.
The window-level models are trained to directly optimize per-window focal loss. 

We use the Adam optimizer \cite{kingma2017adam} to train all models. After tuning on the validation set, we selected a learning rate of $10^{-7}$ for the proposal-based and $10^{-4}$ for the window-level models.
We train each model until validation loss stops improving, evaluating every 3000 training episodes on a random sample of 500 validation set episodes.

\section{Experiments and Results}

Here we report experimental conditions and results on the ESC-CASE and Vox-CASE benchmarks.

\begin{table*}
\centering
\begin{tabular}{|lrrrrrr|}
\hline
\multicolumn{7}{|c|}{\textbf{ESC-CASE}}  \\ \hline
\multicolumn{1}{|c|}{} &  \multicolumn{3}{c|}{5-shot} & \multicolumn{3}{c|}{10-shot}  \\ \hline
\multicolumn{1}{|c|}{Model}              & \multicolumn{1}{r}{AP} & \multicolumn{1}{r}{Acc.} & \multicolumn{1}{r|}{F1} & \multicolumn{1}{r}{AP} & \multicolumn{1}{r}{Acc.} & \multicolumn{1}{r|}{F1} \\ \hline
\multicolumn{1}{|l|}
{Window-Level CRNN} & 0.545 $\pm$ .003 & 0.855 $\pm$ .001 & \multicolumn{1}{r|}{0.280} 
                    & 0.525 $\pm$ .003 & 0.850 $\pm$ .001 &  0.263  \\
\multicolumn{1}{|l|}{Proto-RCRNN}       &  
{\bf 0.935} $\pm$ .002 & 0.937 $\pm$ .001 & \multicolumn{1}{r|}{{0.446}} & 
{0.917} $\pm$ .002 & {\bf 0.943} $\pm$ .001 & 0.471  \\ 
\multicolumn{1}{|l|}{Proto-RCRNN + Perceiver}       & 0.922 $\pm$ .002 & {\bf 0.940} $\pm$ .001 & \multicolumn{1}{r|}{\bf 0.483} & {\bf 0.924} $\pm$ .002 & 0.940 $\pm$ .001 &  \textbf{0.474}  \\
\hline
\end{tabular}
\\
\caption{\label{tab:esccase}5-way 5-shot and 10-shot experiment results for ESC-CASE using both the window-level baseline and proposal-based models. Average Precision (AP) is reported for the proposal-based models after the refinement. Best results are in bold.}
\end{table*}

\begin{table*}
\centering
\begin{tabular}{|lrrrrrr|}
\hline
\multicolumn{7}{|c|}{\textbf{Vox-CASE}}  \\ \hline
\multicolumn{1}{|c|}{} &  \multicolumn{3}{c|}{5-shot} & \multicolumn{3}{c|}{10-shot}  \\ \hline
\multicolumn{1}{|c|}{Model}              & \multicolumn{1}{r}{AP} & \multicolumn{1}{r}{Acc.} & \multicolumn{1}{r|}{F1} & \multicolumn{1}{r}{AP} & \multicolumn{1}{r}{Acc.} & \multicolumn{1}{r|}{F1} \\ \hline
\multicolumn{1}{|l|}
{Window-Level CRNN} & 0.792 $\pm$ .003 & 0.885 $\pm$ .001 & \multicolumn{1}{r|}{0.376} 
                    & 0.815 $\pm$ .002 & 0.897 $\pm$ .001 &  0.407  \\
\multicolumn{1}{|l|}{Proto-RCRNN}       &  
{0.942} $\pm$ .002 & {\bf 0.934} $\pm$ .001 & \multicolumn{1}{r|}{{0.391}} & 
{0.958} $\pm$ .001 & {\bf 0.944} $\pm$ .001 & \textbf{0.465}  \\ 
\multicolumn{1}{|l|}{Proto-RCRNN + Perceiver}       & {\bf 0.953} $\pm$ .001 & {\bf 0.932} $\pm$ .001 & \multicolumn{1}{r|}{\bf 0.418} & {\bf 0.958} $\pm$ .002 & 0.939 $\pm$ .001 &  0.464  \\
\hline
\end{tabular}
\\
\caption{\label{tab:voxcase}5-way 5-shot and 10-shot experiment results for Vox-CASE using both the window-level baseline and proposal-based models. Average Precision (AP) is reported for the proposal-based models after the refinement. Best results are in bold.}
\end{table*}

\begin{table*}[]
\centering
\begin{tabular}{|l|llll|} \hline
  & ESC-CASE 5-shot & ESC-CASE 10-shot & Vox-CASE 5-shot & Vox-CASE 10-shot \\ \hline
 Proto-RCRNN & 0.623 
 / { 0.936} 
 & {0.562} 
 / 0.917
 & {0.688} 
 / 0.942
 & 0.713 
 / 0.958 
 \\ 
+ Perceiver & 0.656 / 0.923 
 & 0.653 / 0.924
 & 0.748 / 0.953 
 & 0.761 / 0.958 \\ \hline
\end{tabular}
\\
 \caption{\label{tab:ap}Average Precision (AP) is reported after stage I and stage II for the proposal models on ESC-CASE and Vox-CASE.}
\end{table*}

\subsection{Metrics}
For all models, we report three metrics:
\subsubsection{Average precision (AP)} For the event / no event binary classification task (using the second stage for the proposal-based models). AP measures how well we localize events (vs background), agnostic to any particular class. This is computed per test set episode, and both the mean and standard deviation over test set episodes are reported.
\subsubsection{Accuracy} Window- and proposal-level accuracy are computed per episode, and both the mean and standard deviation over test set episodes are reported. Accuracy tells us how good the model is at assigning the correct class label (or background) to each window (window-level) or proposal (proposal-based). 
\subsubsection{F1 score}
F1 is computed over proposals or proposal-like (see below) units, Accordingly, it assesses both localization and classification.  
F1 is computed per class over all episodes, not per episode, and is then averaged over classes. Given the relatively fewer classes to average, only mean is reported.

For the sake of consistency in evaluating the window-level and proposal-based models, the window-level's predictions are mapped into proposal-like regions and then evaluated at this ``proposal'' level. Specifically, for each window, we choose the most probable class, and then group consecutive windows with the same class prediction. 
This yields the proposal-like regions with their class labels, which we translate into proposal timestamps in seconds. Finally, we associate confidence scores with each proposal by averaging the window-level posterior probabilities over each proposal.

\subsection{Results and Analysis}\label{sec:results}
Our main set of results for ESC-CASE are reported in Table~\ref{tab:esccase}. First, we note that the proposal models outperform the window-level model. Comparing 5-shot and 10-shot results, there is similar performance according to AP, accuracy, and F1, suggesting that additional training examples per class may not offer much benefit.
Overall, the use of Perceivers improves performance, with the Proto-RCRNN + Perceiver performing the best.

Table~\ref{tab:voxcase} presents the same set of comparisons on the Vox-CASE dataset, where the proposal-based models continue to perform well via AP and accuracy, but the window-level model offers surprisingly strong F1. The Perceiver module does not increase performance for Vox-CASE to the same degree that it does for ESC-CASE.

To dig into the effect of the second refinements for the proposal-based models, we report AP after the RPN (left of slash), and after the second set of refinements (right of slash) in Table \ref{tab:ap}. We note that AP after the refinement layer is always significantly higher than AP directly from the RPN, 
suggesting that the refinement layer is necessary to achieve accurate localization of sound events.

\begin{figure*}
  \includegraphics[width=0.5\linewidth]{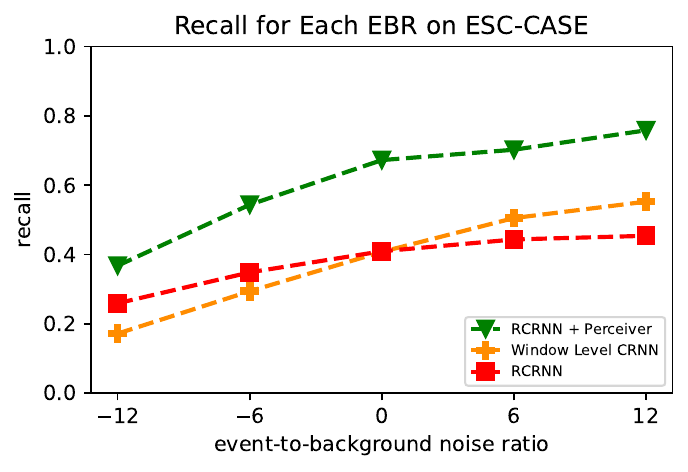}
  \includegraphics[width=0.5\linewidth]{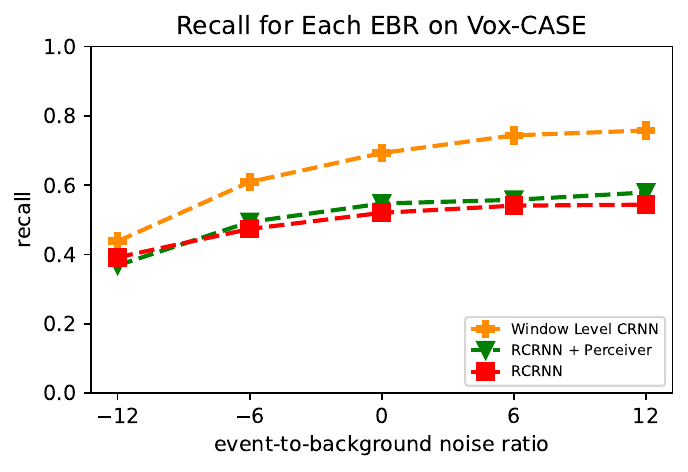}
  \caption{F1 score vs event-to-background radio for ESC-CASE (left) and Vox-CASE (right).}
  \label{fig:ebr1}
\end{figure*}

As described in Section \ref{sec:datagen}, our events vary in their energy relative to the background noise. Figure \ref{fig:ebr1} shows event recall as a function of EBR for both the window-level and proposal models on ESC-CASE (left) and Vox-CASE (right). 
We report recall to emphasize the relative ability to detect events.
As expected, performance increases as the events become louder relative to the background. For ESC-CASE, the proposal-based model with the Perceiver module seems to perform the best based on the recall metric, while the proposal-based model without the Perceiver performs similarly to our window-level baseline. For Vox-CASE, the window-level baseline is highest at all EBRs, while the proposal-based models seem to perform identically. We hypothesize that the Perceiver may assist the proposal-based model in differentiating between background and events for the ESC-CASE dataset, where there is greater acoustic similarity between events and backgrounds. For the Vox-CASE dataset, the Perceiver may provide less of a benefit because events and backgrounds are more acoustically dissimilar. When events and backgrounds are already acoustically dissimilar, it is easier for the proposal-based model to distinguish between them based on their acoustic features. Therefore, adding an additional deep learning module, such as the Perceiver, may not significantly improve the model's performance.

\section{Conclusion and Future Work}
In this work, we presented a proposal-based approach to few-shot sound event detection, and introduced a Perceiver variant of the proposal-based method. We evaluated the models on two new datasets designed for episodic few-shot training and evaluation: ESC-CASE with environmental sounds as events and Vox-CASE with celebrity voices as events.

For both ESC-CASE and Vox-CASE, we find consistently better performance with the proposal-based model compared to a strong window-level baseline. However, improvement is substantially larger for ESC-CASE than for Vox-CASE.
This suggests the proposal model may be superior in cases where the events and background are more similar.
We conclude that proposal-based methods, already shown to be successful in the non-few-shot case and for few-shot video action recognition, have an important role to play for few-shot sound event detection, and that the Perceiver model can further improve performance when used in place of RoI pooling.

Prototypical networks are popular and effective, but alternative few-shot learners may yield even better performance. Lastly, pushing the limits of the task with lower $k$ (shot) or higher $n$ (way) may reveal different trade-offs between the proposal- and window-based approaches.

\bibliographystyle{IEEEtran}
\bibliography{refs}

\end{document}